\def\CMP{\sevenrm Commun.\ Math.\ Phys.}
\def\JMP{\sevenrm J.\ Math.\ Phys.}

\def\RMP{\sevenrm Rev.\ Math.\ Phys.}
%
%
\def\today{\number\day .\space\ifcase\month\or
January\or February\or March\or April\or May\or June\or
July\or August\or September\or October\or November\or December\fi, \number \year}
%
%
\newcount \theoremnumber
\def\cleartheoremnumber{\theoremnumber = 0 \relax}

\def\Prop #1 {
             \advance \theoremnumber by 1
             \vskip .6cm 
             \goodbreak 
             \noindent
             {\bf Propsition {\the\headlinenumber}.{\the\theoremnumber}.}
             {\sl #1}  \goodbreak \vskip.8cm}

\def\Conj#1 {
             \advance \theoremnumber by 1
             \vskip .6cm  
             \goodbreak 
             \noindent
             {\bf Conjecture {\the\headlinenumber}.{\the\theoremnumber}.}
             {\sl #1}  \goodbreak \vskip.8cm} 

\def\Th#1 {
             \advance \theoremnumber by 1
             \vskip .6cm  
             \goodbreak 
             \noindent
             {\bf Theorem {\the\headlinenumber}.{\the\theoremnumber}.}
             {\sl #1}  \goodbreak \vskip.8cm}

\def\Lm#1 {
             \advance \theoremnumber by 1
             \vskip .6cm  
             \goodbreak 
             \noindent
             {\bf Lemma {\the\headlinenumber}.{\the\theoremnumber}.}
             {\sl #1}  \goodbreak \vskip.8cm}

\def\Cor#1 {
             \advance \theoremnumber by 1
             \vskip .6cm  
             \goodbreak 
             \noindent
             {\bf Corollary {\the\headlinenumber}.{\the\theoremnumber}.}
             {\sl #1}  \goodbreak \vskip.8cm} 
%
%
\newcount \equationnumber

\newcount \refnumber

\def\[]    {\global 
            \advance \refnumber by 1
            [{\the\refnumber}]}

\def\# #1  {\global 
            \advance \equationnumber by 1
            $$ #1 \eqno ({\the\equationnumber}) $$ }

\def\% #1 { \global
            \advance \equationnumber by 1
            $$ \displaylines{ #1 \hfill \llap ({\the\equationnumber}) \cr}$$} 

\def\& #1 { \global
            \advance \equationnumber by 1
            $$ \eqalignno{ #1 & ({\the\equationnumber}) \cr}$$}
%
%
\newcount \Refnumber

\def\Ref #1 #2 #3 #4 #5 #6  {\ninerm \global
                             \advance \Refnumber by 1
                             {\ninerm #1,} 
                             {\ninesl #2,} 
                             {\ninerm #3.} 
                             {\ninebf #4,} 
                             {\ninerm #5,} 
                             {\ninerm (#6)}\nobreak} 
\def\Bookk #1 #2 #3 #4       {\ninerm \global
                             \advance \Refnumber by 1
                             {\ninerm #1,}
                             {\ninesl #2,} 
                             {\ninerm #3,} 
                             {(#4)}}
\def\Book{\cr
{\the\Refnumber} &
\Bookk}
\def\Reff{\cr
{\the\Refnumber} &
\Ref}
\def\REF #1 #2 #3 #4 #5 #6 #7   {{\sevenbf [#1]}  & \hskip -9.5cm \vtop {
                                {\sevenrm #2,} 
                                {\sevensl #3,} 
                                {\sevenrm #4} 
                                {\sevenbf #5,} 
                                {\sevenrm #6} 
                                {\sevenrm (#7)}}\cr}
\def\BOOK #1 #2 #3 #4  #5   {{\sevenbf [#1]}  & \hskip -9.5cm \vtop {
                             {\sevenrm #2,}
                             {\sevensl #3,} 
                             {\sevenrm #4,} 
                             {\sevenrm #5.}}\cr}
\def\HEP #1 #2 #3 #4     {{\sevenbf [#1]}  & \hskip -9.5cm \vtop {
                             {\sevenrm #2,}
                             {\sevensl #3,} 
                             {\sevenrm #4.}}\cr}
%
%
\def\bull{$\sqcup \kern -0.645em \sqcap$}
%
%
\def\Def#1{\vskip .3cm \goodbreak \noindent
                                     {\bf Definition.} #1 \goodbreak \vskip.4cm}
\def\Rem#1{\vskip .4cm \goodbreak \noindent
                                     {\it Remark.} #1 \goodbreak \vskip.5cm }

\def\Pr#1{\goodbreak \noindent {\it Proof.} #1 \hfill \bull  \goodbreak \vskip.5cm}

%
%
\def\*{\vskip 1.0cm}      

%
%
\newcount \ssubheadlinenumber

\def\SSHL #1 {\goodbreak
            \cleartheoremnumber
            \vskip 1cm
            \advance \ssubheadlinenumber by 1
{\rm \noindent {\the\headlinenumber}.{\the\subheadlinenumber}.{\the\ssubheadlinenumber}. #1}
            \nobreak \vskip.8cm \rm \noindent}
\newcount \subheadlinenumber
\def\clearsubheadlinenumber{\subheadlinenumber = 0 \relax}
\def\SHL #1 {\goodbreak
            \cleartheoremnumber
            \vskip 1cm
            \advance \subheadlinenumber by 1
            {\rm \noindent {\the\headlinenumber}.{\the\subheadlinenumber}. #1}
            \nobreak \vskip.8cm \rm \noindent}
\newcount \headlinenumber

\newcount \headlinesubnumber
\def\clearheadlinesubnumber{\headlinesubnumber = 0 \relax}
\def\Hl #1 {\goodbreak
            \cleartheoremnumber
            \clearheadlinesubnumber
            \clearsubheadlinenumber
            \advance \headlinenumber by 1
            {\bf \noindent {\the\headlinenumber}. #1}
            \nobreak \vskip.4cm \rm \noindent}

\font\twentyrm=cmr17
\font\fourteenrm=cmr10 at 14pt
\font\sevensl=cmsl10 at 7pt
\font\sevenit=cmti7 

\font\css=cmss10
\font\Rosch=cmr10 at 9.85pt
\font\Cosch=cmss12 at 9.5pt
\font\rosch=cmr10 at 7.00pt
\font\cosch=cmss12 at 7.00pt
\font\nosch=cmr10 at 7.00pt
%
%
%
%
%
%
%
%
%
%
%
%
%
%
%
%
\def\Z                 {\hbox{{\css Z}  \kern -1.1em {\css Z} \kern -.2em }}
\def\R                 {\hbox{\raise .03ex \hbox{\Rosch I} \kern -.55em {\rm R}}}
\def\N                 {\hbox{\rm I \kern -.55em N}}
\def\C                 {\hbox{\kern .20em \raise .03ex \hbox{\Cosch I} \kern -.80em {\rm C}}}

\def\r                 {\hbox{\raise .03ex \hbox{\rosch I} \kern -.45em \hbox{\rosch R}}}
\def\n                 {\hbox{\hbox{\rosch I} \kern -.45em \hbox{\nosch N}}}
\def\c                 {\hbox{\raise .03ex \hbox{\cosch I} \kern -.70em \hbox{\rosch C}}}

\def\z                 {\hbox{\kern 0.2em {\cal z}  \kern -0.6em {\cal z} \kern -0.3em  }}
\def\1                 {\hbox{\rm \thinspace \thinspace \thinspace \thinspace
                                  \kern -.50em  l \kern -.85em 1}}
\def\unit                 {\hbox{\sevenrm \thinspace \thinspace \thinspace \thinspace
                                  \kern -.50em  l \kern -.85em 1}}
%
%
%
%
%
%
%
%
%
%
%
%
\def\Tr                {{\rm Tr}}
\def\A                 {{\cal A}} 
 
\def\D                 {{\cal D}} 
 
\def\B                 {{\cal B}} 

\def\H                 {{\cal H}} 
 
\def\O                 {{\cal O}}

%

%
%
%
%
%
%
%


\def\versuch #1 #2 {
\vskip -.1 cm
\global \advance \equationnumber by 1
            $$\displaylines{ \rlap{ #1 } \hfill #2  \hfill \llap{({\the\equationnumber})} } $$ 
\vskip  .1cm
\noindent}

\nopagenumbers
\def\Draft  {\hbox{Preprint \today}}
\def\firstheadline{\hss \hfill  \Draft  \hss} 
\headline={
\ifnum\pageno=1 \firstheadline
\else 
\ifodd\pageno \rightheadline 
\else \leftheadline \fi \fi}
\def\rightheadline{\sevenrm TWO ALGEBRAIC PPROPERTIES OF THERMAL QUANTUM FIELD THEORIES  
\hfill \folio } 
\def\leftheadline{\sevenrm \folio \hfill CHRISTIAN D.\ J\"AKEL}
\voffset=2\baselineskip
\magnification=1200
%
%
%
%
%

\vskip 1cm

\noindent
{\twentyrm Two algebraic properties of thermal}
                  
\vskip .2cm
\noindent
{\twentyrm quantum field theories}
                  
\vskip 1cm
                  
\noindent
{\sevenrm Christian D.\ J\"akel}\footnote{$^{a)}$}
{\sevenrm Present address: Institut f.\
theor.\ Physik, Universit\"at Innsbruck, Austria, 
\hfill
\break \indent
Electronic mail:
christian.jaekel@uibk.ac.at}

\noindent
{\sevenit  Dipartimento di Matematica,
Via della Ricerca Scientifica, Universit\`a di Roma ``Tor Vergata'', 
I-00133~Roma, e-mail: cjaekel@esi.ac.at}

\vskip .5cm     
\noindent {\sevenbf Abstract}. {\sevenrm We establish 
the Schlieder and the Borchers property for thermal field theories.
In addition, we provide some information on the commutation and localization
properties of projection operators.}

\vskip 1 cm

%
%
%
%
%

\Hl{Introduction}

\noindent
Recently, the author has explored the general structure of
thermal field theories in some detail. Here we would like
to fill in a gap which concerns two basic results, namely, the Borchers and the 
Schlieder property. Both results will look familiar to the experts and even 
the proofs which we will present here 
are more or less standard (see, e.g., Ref.\ 1 for a convenient collection of 
the corresponding results in the vacuum sector). However, a close inspection shows that  
the fundamental differences between thermal and vacuum QFT are clearly reflected in  
slightly different assumptions and consequences. For completeness we add a
list of properties (due to Florig and Summers$^2$) which are all equivalent to 
the Schlieder property. 

In the algebraic formulation (as described in the monograph by Haag$^3$) a QFT is 
cast into an inclusion preserving map 
\# {\O \to \A (\O)} 
which assigns to any open bounded region $\O$ in Minkowski 
space $\R^4$ a unital $C^*$-algebra $\A (\O)$. 
The Hermitian elements of the {\sl abstract} $C^*$-algebra $\A (\O)$ are interpreted as 
the observables which can be measured at times and locations in $\O$. 
The physical states are described by positive, linear, and normalized functionals. 
By the GNS-construction, any state $\omega$ on $\A$ gives rise to a 
Hilbert space~$\H_\omega$ and a representation $\pi_\omega$
together with a cyclic vector~$\Omega_\omega$, 
such that
\# { \omega (a) = \bigl( \Omega_\omega \, , \, \pi_\omega (a) \Omega_\omega \bigr) 
\qquad \forall a \in \A = 
\overline{ \cup_{ {\cal O}  \subset \r^4} \A(\O) }^{\, C^*} . }
The representation $\pi_\omega$ 
automatically determines the values of certain macroscopic observables  
in all states, which are normal (A linear functional on
${\cal A}$ is said to 
be normal relative to $\pi_\omega$, if it is continuous with respect to 
the ultraweak topology determined by $\pi_\omega$.
Since normal states differ only locally from $\omega$,  
various global physical situations will manifest themselves in
unitarily inequivalent GNS-representations.)
w.r.t.\ $\pi_\omega$ (these are exactly those states which
can be specified by density matrices $\rho \in \B(\H_\omega)$, 
$\rho >0$, $\Tr \, \rho = 1$). Thus a state is specified macroscopically by a 
representation and microscopically by a density matrix.

The relevant states describing thermal equilibrium, the so-called KMS-states, 
will soon be distinguished within the set of all time-invariant normalized, positive linear
functionals of~$\A$ by their stability properties with respect to timelike translations.
Since the associated GNS-representations will allow a unitary implementation of the 
time-evolution, we will take advantage of it and formulate our problems in the better 
developed Hilbert space setting.
But before we do so, we should mention that the net $\O \to \A(\O)$ satiesfies a number of 
properties which do not depend on the representation:

\vskip .2cm 
\noindent
\halign{ #  \hfil & \vtop { \parindent =0pt \hsize=36,6em
                            \strut # \strut} \cr 
(i)  & The net $\O \to \A(\O)$ is isotonous, i.e., there exists a unital embedding
\# {\A(\O_1) \hookrightarrow \A(\O_2)
\qquad \hbox{if} \quad \O_1 \subset \O_2.} 
This property, called isotony, allows us to consider the quasilocal algebra
$\A$ which is defined in (2) as the $C^*$-inductive limit of the local algebras.
The elements of $\A$ are called quasilocal observables; they can be approximated in norm 
topology by strictly local elements; the total energy, total charge, etc., 
are considered as unobservable; these quantities refer to infinitely extended regions
and can not be controlled by local measurements.
\cr
(ii)   & Observables localized in spacelike separated space--time regions commute,
\#
{\A (\O_1) \subset \A^c ( \O_2) \quad \hbox{\rm if} \quad \O_1 \subset \O_2'.}
Here $\O '$ denotes the spacelike complement of $\O$ and 
$\A^c (\O)$ denotes the set of operators in~$\A$ which commute with all operators in $\A(\O)$.
\cr
(iii)    & The space--time symmetry of Minkowski space manifests itself
in the existence of a representation 
\# {\alpha \colon ( \Lambda, x) \mapsto \alpha_{ \Lambda, x} \in \hbox{Aut} (\A) ,
\qquad (\Lambda, x) \in {\cal P}_+^\uparrow}, 
of the (orthochronous)
Poincar\'e group ${\cal P}_+^\uparrow$. Lorentz-transformations $\Lambda$  and space--time 
translations~$x$ act geometrically,
\# {\alpha_{ \Lambda, x} \bigl( \A (\O) \bigr) 
= \A (\Lambda \O + x)  \qquad \forall (\Lambda, x) \in {\cal P}_+^\uparrow.} 
For the present letter we may restrict our list of assumptions to include only 
the (strongly continuous) one-parameter subgroup of time-translations 
$\tau \colon \R \to \hbox{Aut}(\A)$. Of course, it acts geometrically, i.e.,
\# {\tau_t \bigl( \A (\O) \bigr) 
= \A (\O +te)  \qquad \forall t \in \R.} 
Here $e$ is a unit vector denoting the time 
direction with respect to a given Lorentz-frame. 
\cr}

\Rem{Let $h \in L^1 (\R , {\rm d} t)$  such that the Fourier-transform
$\tilde h$ of $h$ has compact support. If the group of
automorphisms  $\tau \colon \R \to \hbox{Aut}(\A)$ is strongly continuous, then the Bochner integral 
\# {a_h = \int {\rm d} t \, h(x) \tau_t (a) ,  \qquad a \in \A , }
exists in $\A$ and defines an entire analytic element for the time-translations. 
Recall that
$b \in \A$ is called an analytic element for the group~$\tau$, if
there exists some $\lambda > 0$, a strip $S(_ \lambda , \lambda) := \{ z \in \C :
| \Im z | < \lambda \}$ and a function
$g \colon {\cal N} \to \A$ such that
\vskip .2cm
\hskip 1cm
(a) $\qquad g(t) = \tau_t (b)$; 
\vskip .2cm
\hskip 1cm
(b) $\qquad z \mapsto \omega \bigl( g(z) \bigr)$ 
is analytic for all states $\omega$ over ${\cal A}$.
\vskip .2cm
\noindent
An element is called entire analyitc if  $\lambda = \infty$; i.e.,
$S(_ \lambda , \lambda) = \C$
$\A_\tau$ will denote the set of entire analytic elements.}

We now turn to thermal equilibrium states. Kubo$^4$ and 
subsequently Martin and Schwinger$^5$ found out that the Green's functions of
finite-volume Gibbs-states satisfy an auxiliary boundary condition
in the complex plane with respect to the time-evolution.
The crucial step was to recognize that the so-called KMS-condition not only characterizes the
finite-volume Gibbs-states but remains valid in the thermodynamic limit$^6$.
Nowadays, the KMS-condition is accepted as {\sl the} appropriate
criterion for equilibrium of 
finite {\sl and} infinite systems. But only recently, Buchholz and Junglas have shown that a 
large class of relativistic models admits KMS-states$^7$.    

\Def{A state
$\omega_\beta$ over $\A$ is called a $(\tau , \beta )$-KMS-state for 
some $\beta \in \R \cup \{ \pm \infty \}$, if 
\# { \omega_\beta \bigl( a \tau_{i \beta} (b) \bigr) = \omega_\beta (b a) }
for all $a, b$ in a norm dense,
$\tau$-invariant, $*$-subalgebra of $\A_\tau$.}

The GNS-representation $(\pi_\beta, \H_\beta, \Omega_\beta)$ of the net $\O \to \A(\O)$
associated with a KMS-state~$\omega_\beta$ 
assigns to any $\O \subset \R^4$ a von Neumann algebra
\# { \qquad {\cal R}_\beta (\O) = \pi_\beta \bigl( \A(\O) \bigr)''. }
${\cal R}_\beta := \pi_\beta (\A)''$ possesses a cyclic 
(due to the GNS-construction) and separating (due to the KMS-condition) 
vector, namely, $\Omega_\beta$.

The general analysis of KMS-states (see, e.g., Ref.\ 8) extends a number
of results well known in classical ergodic theory to the 
noncommutative case. For instance, the set of KMS-states for any fixed 
$\beta > 0$ is a weak-* compact, convex set. In fact, an arbitrary KMS-state 
can be represented 
in a unique manner as a convex superposition of extremal 
KMS-states. (A KMS-state is called extremal, if it cannot 
be decomposed into other KMS-states.) Moreover, KMS-states can be distinguished 
within the set of all (physical) states from first principles.
In a number of pioneering articles it has been demonstrated that
the extremal KMS-states of an infinitely extended medium change 
continuously as the Hamiltonian is perturbed slightly. This condition characterizes 
the extremal KMS-states; they are precisely those states which are
distinguished among (possible other) stationary states by the fact that they 
turn continuously into the unperturbed states as a certain family of perturbations tends to
zero.$^{9,10}$  The same condition
may also be interpreted as an adiabatic invariance.$^{11}$ 
Extremal KMS-states return to their original form at the end of 
a procedure in which the dynamical law is changed by a local perturbation which is slowly
switched on and, as $t \to \infty$ slowly switched off again. 
A second important characteristic of KMS-states is their passivity,$^{12}$ 
which is the requirement 
that the energy of the system at time $t$ can only have increased if the Hamiltonian depends on the
time and has returned to its initial form at time $t$. 
This condition is just the second law of thermodynamics; it fixes the sign of 
$\beta$ and means that no energy can be removed from a KMS-state having $\beta >0$, 
just as a periodic process can extract no energy from the ground state.

Although the representation independent aspects of the map $\O \to \A(\O)$
clearly deserve attention [it seems that the abstract 
operator algebraic formulation is inevitable for the describtion of nonequilibrium 
situations in which 
also the macroscopic observables (e.g., the specific heat, the mean magnetization, etc.) 
will change in the course of time], we will now specify our results and assumptions 
in the more restrictive Hilbert space framework.
To be precise, in the remainder of this letter we will consider a TFT, specified by a
von Neumann algebra ${\cal R}_\beta$ with a cyclic and separating
vector $\Omega_\beta$ and a net of subalgebras
\# { \O \to {\cal R}_\beta (\O)  } 
subject to the following conditions: 
\vskip .3cm
\noindent
\halign{ #  \hfil & \vtop { \parindent =0pt \hsize=36,6em
                            \strut # \strut} \cr 
(i)     & The subalgebras associated with spacelike 
separated space--time regions commute, i.e.,
\# { {\cal R}_\beta (\O_1) \subset {\cal R}_\beta (\O_2)'
\qquad \hbox {if} \qquad \O_1 \subset \O_2'.}
Note that ${\cal R}_\beta ( \O)'$ denotes the commutant of 
${\cal R}_\beta ( \O)$ in $\B(\H_\beta)$, an algebra that includes both ${\cal R}_\beta'$ 
and ${\cal R}_\beta (\O')$ as subalgebras. 
\cr
(ii)     & The time-evolution is unitarily implemented by the modular
group
$t \mapsto \Delta^{it}$ (see, e.g., Ref.\ 8) associated with 
the pair $({\cal R}_\beta, \Omega_\beta)$, i.e.,
\# {  \Delta = {\rm e}^{- \beta H_\beta} \quad \hbox{and} \quad 
\Delta^{-i t / \beta} {\cal R}_\beta (\O)   \Delta^{i t / \beta} 
= {\cal R}_\beta (\O + t e) 
\qquad 
\forall t \in \R.}
Here $H_\beta$ denotes the generator of the time-evolution and 
$e$ is the unit vector denoting the time 
direction w.r.t.\  the distinguished  rest frame. 
\cr
(iii)     & $\H_\beta$ is separable and $\Omega_\beta$ is the unique 
           --- up to a phase --- time-invariant vector in~$\H_\beta$. 
\cr
iv.)    &  $\Omega_\beta$ is cyclic for ${\cal R}_\beta (\O)$,
           where $\O$ is any open subset of~$\R^4$,
           i.e., 
\# { \overline { {\cal R}_\beta (\O) \Omega_\beta} = \H_\beta. }
$\Omega_\beta$  
shares the ``Reeh-Schlieder property'' (14)  with the following dense set of 
vectors: 
\# {  {\cal D}_\tau = \{ \Psi = \pi_\beta (a) \Omega_\beta \in \H_\beta : 
a, a^{-1} \in \A_\tau \} \subset {\cal R}_\beta \Omega_\beta .}
\cr}
\vskip .3cm
\noindent
We will show that under these assumptions the following statements are valid: 
\vskip .3cm
\noindent
\halign{ #  \hfil & \vtop { \parindent =0pt \hsize=36,6em
                            \strut # \strut} \cr 
(i)     & (Schlieder property). 
Given two open space--time regions $\O$, $\hat {\O}$ and some $\delta > 0$ such that
\# { \O + te  \subset \hat {\O}  \quad \hbox{for} \quad |t| < \delta ,}
the Schlieder property holds for 
the algebras ${\cal R}_\beta (\O)$ and ${\cal R}_\beta (\hat {\O})'$.
(Recall that a pair of von Neumann algebras 
${\cal M}$ and ${\cal N}$ satisfies the Schlieder property iff $0 \ne M \in {\cal M}$
and $0 \ne N \in {\cal N}$ implies that $MN \ne 0$.) 
\cr
(ii)     & 
(Borchers property; also called Property B). 
Given a nonzero projection operator $E \in {\cal R}_\beta(\O)$, where 
$\O \subset \R^4$ is bounded,
there exists a partial isometry~$V$ in the von Neumann algebra ${\cal R}_\beta(\hat{\O})$,
corresponding to a slightly larger region $\hat \O$, such that $V^*V = \1 $ and $V V^* = E$.
One writes  
\# { E \sim \1 _{\hbox{\fiverm mod} \, {\cal R}_\beta (\hat {\cal O}) } }
Recall that a factor ${\cal M}$ is called type~III, if 
$E \sim \1 _{\hbox{\fiverm mod} \, {\cal M} } $ for all self-adjoint
projections $E \in {\cal M}$. Thus ${\cal R}_\beta (\O)$ is ``almost''
a factor of type~$III$. 
\cr}

\Rem{If the pair $(\A, \tau)$ is asymptotically Abelian in time, i.e.,
\# { \lim_{t \to \infty} \bigl\| [ a, \tau_t (b) ] \bigr\| = 0 \qquad \forall a, b \in \A }
and
$\omega_\beta$ is an extremal KMS state, then
$\Omega_\beta$ is the unique --- up to a phase --- time-invariant vector in $\H_\beta$. 
The Reeh--Schlieder property (iv)\ can be derived 
from the relativistic KMS-condition$^{13}$ provided the net $ \O \to {\cal R}_\beta (\O)$
satisfies additivity.{$^{14}$} [The net $\O \to {\cal R}_\beta (\O)$ is called additive  if
$\cup_i \O_i = \O \Rightarrow \vee_i {\cal R}_\beta (\O_i) 
= {\cal R}_\beta (\O)  $. Here ${\cal R}_1 \vee {\cal R}_2$ denotes the von Neumann
algebra generated by the algebras ${\cal R}_1 $ and ${\cal R}_2$.] Junglas$^{15}$ 
has shown that the Reeh-Schlieder 
property of $\Omega_\beta$ also follows from the 
standard KMS-condition, if $\omega_\beta$ is locally normal 
w.r.t.\ the vacuum representation.
Note that we do {\sl not} require that there exists a group of 
unitary operators in $\B(\H_\beta)$ 
which implements spacelike translations, since spatial translation invariance may be 
spontaneously broken in a KMS-state. 
}

\vskip 1cm

\Hl{Two Basic Properties of TFTs}

\noindent
We start with the analogon of a result of Borchers (see Ref.\ 16).

\Lm{Let $E \in {\cal R}_\beta$, $\| E \| =1$ and let
$F = F^* = F^2 \in \B(\H_\beta)$ be a projection operator such that
\# { \bigl[ {\rm e}^{- it  H_\beta} F {\rm e}^{it  H_\beta} \, , \, E \bigr] = 0
\qquad \forall |t| < \delta .}
Then  $FE  = 0$ implies
$ F {\rm e}^{ it  H_\beta} E = 0$  for all $t \in \R $.}

\Pr{It is sufficient to show that 
\# { \bigl( \Phi \, , \, {\rm e}^{ - it  H_\beta} 
F {\rm e}^{ it  H_\beta} E \Psi \bigr) = 0 
\qquad \forall t \in \R }
for the dense set of vectors $\Phi, \Psi \in {\cal D}_\tau$ introduced in (15). 
By definition, the vectors in ${\cal D}_\tau$ are entire analytic for the energy, i.e.,
${\cal D}_\tau \subset  \D \bigl( {\rm e}^{ - z  H_\beta} \bigr)$
for all $ z \in \C $.
Due to the KMS-relation,
\# { {\cal R}_\beta \Omega_\beta \subset  \D \bigl( {\rm e}^{ - \lambda  H_\beta} \bigr) \qquad
\forall 0 \le \lambda \le  \beta / 2 .}
Thus the function
\# { z \mapsto
\bigl(  {\rm e}^{ i \bar z  H_\beta} \Phi \, , \,  
F {\rm e}^{ iz  H_\beta}  E \Psi \bigr) }
is analytic in the strip $0 < \Im z < \beta / 2$, while the function
\# { z \mapsto
\bigl( {\rm e}^{  i \bar z H_\beta} E^* \Phi \, , \,  F 
{\rm e}^{  i z H_\beta} \Psi \bigr) }
is analytic in the strip $- \beta / 2 < \Im z < 0$. 
Both functions are bounded and analytic
and have continuous 
boundary values for $\Im z \searrow 0$ and $\Im z \nearrow 0$, respectively.
Now (19) implies
\# { 
\lim_{\Im z \searrow 0}  \bigl(  {\rm e}^{ i \bar z  H_\beta} \Phi \, , \,  
F {\rm e}^{ iz  H_\beta}  E \Psi \bigr) 
= \lim_{\Im z \nearrow 0}  \bigl( {\rm e}^{  i \bar z H_\beta} E^* \Phi \, , \,  F 
{\rm e}^{  i z H_\beta} \Psi \bigr)
\qquad  \forall |t| < \delta.}
Using the Edge-of-the-Wedge Theorem$^{17}$ one concludes that there exists a function 
\# { f_{E,F} \colon G_{\delta} \to \C } 
which is analytic on the doubly cut strip
\# { G_{\delta} = \{ z \in \C : - \beta / 2  < \Im z  < \beta / 2 \} \setminus \{ z \in \C : 
\Im z = 0, |\Re z| \ge \delta \}   }
and satisfies
\# {f_{E,F} (z) = 
\left\{
\eqalign{
&  {  \bigl( {\rm e}^{ i \bar z  H_\beta}
\Phi \, , \,  F {\rm e}^{ i z  H_\beta} E \Psi \bigr) }
\cr
&  {  \bigl( {\rm e}^{  i \bar z H_\beta} E^*  \Phi \, , \,  F 
{\rm e}^{  i z H_\beta} \Psi \bigr) }}
\right\} 
{\rm \ for \ }  
\left\{
\eqalign{
& { 0  < \Im z   <  \beta / 2 ,}  
\cr
& { - \beta /2  < \Im z  <  0 .}}  
\right\} 
}
Continuity and $FE = 0$ imply $f_{E,F} (0) = 0$.
According to Lagrange's theorem $f_{E,F} (z)$ vanishes identically 
if $0$ is a zero of infinite order. This follows from the original 
arguments of Borchers; put
\# { t^{(i)}_j := {\delta j \over 2 i n} , \qquad i \in \N, \quad j = \{ 1, \ldots , n \} .}
Now set
\# {
f_{t^{(i)}_1, \ldots, t^{(i)}_n}^+ (z)    := \bigl( {\rm e}^{i \bar z H_\beta} 
\Phi \, , \,  F (t^{(i)}_1)    
\ldots F (t^{(i)}_n) {\rm e}^{iz H_\beta} E \Psi \bigr) \quad \hbox{for}
\quad 0  \le \Im z \le \beta/2  }
and
\# { f_{t^{(i)}_1, \ldots, t^{(i)}_n}^- (z) := 
\bigl( {\rm e}^{ i \bar z  H_\beta} E^* 
\Phi \, , \, 
F (t^{(i)}_1) \ldots F (t^{(i)}_n) {\rm e}^{ i z H_\beta} \Psi \bigr)  \quad \hbox{for}  
\quad 
- \beta /2  \le \Im z  \le  0 , } 
where
\# {F \bigl( t^{(i)}_j \bigr) := {\rm e}^{-i t^{(i)}_j H_\beta} 
F  {\rm e}^{ i t^{(i)}_j H_\beta}. }
Both functions are analytic in the interior of their domains and bounded and continuous at the 
boundary. Since $\bigl| t_j^{(i)} \bigr| \le \delta / 2$ for all  $i \in \N$ and
$j = \{ 1, \ldots , n \}$ implies that
\# { \bigl[  E \, , \,  {\rm e}^{-i t H_\beta}  F (t^{(i)}_1)     
\ldots F (t^{(i)}_n) {\rm e}^{i t H_\beta} \bigr] = 0 \qquad 
\forall | \Re z | < 
\delta / 2 ,}
the boundary values for $ \Im z \searrow 0$ resp.\ 
$\Im z \nearrow 0 $ coincide for $ | \Re z | < 
\delta / 2 $. Applying the Edge-of-the-Wedge Theorem$^{17}$ one concludes
that $f^+$ and $f^-$ are  
the restrictions to the upper (resp.\ lower)
half of the doubly cut strip ${\cal G}_{\delta /2}$ 
of a function  
\# {
f_{t^{(i)}_1, \ldots, t^{(i)}_n} (z) 
:= 
\left\{
\eqalign{
&  { f_{t^{(i)}_1, \ldots, t^{(i)}_n}^+ (z) }
\cr
&
{ f_{t^{(i)}_1, \ldots, t^{(i)}_n}^- (z) } 
}
\right\} 
\hbox{\rm for}
\left\{
\eqalign{
& { 0  < \Im z < \beta/2,}  
\cr
& {-\beta/2 < \Im z < 0 ,}  }  
\right\} 
}
which is defined and analytic for $z \in {\cal G}_{\delta /2}$.
The function 
$ f_{t^{(i)}_1, \ldots, t^{(i)}_n} (z)  $ has continuous boundary values for $z \to \partial 
{\cal G}_{\delta /2}$.
Since $\Phi$ and $\Psi$ are elements of ${\cal D}_\tau$ (see (15)),
there exist two operators $A, B \in \pi_\beta (\A_\tau)$ such that
$\Phi = A \Omega_\beta$ and $\Psi = B \Omega_\beta$.
Applying the maximum modulus 
principle we obtain the following estimate: 
\& {  \bigl| f_{t^{(i)}_1, \ldots, t^{(i)}_n} (z)  \bigr|  
& \le  \sup_{w \in \partial {\cal G}_{\epsilon }} 
\bigl| f_{t^{(i)}_1, \ldots, t^{(i)}_n} (w)  \bigr| 
\cr
& \le \max \Bigl\{ \| {\rm e}^{\beta H_\beta / 2}   A \Omega_\beta \|  
\| B \| \, ,
 \, \| A \|  \| {\rm e}^{ \beta H_\beta / 2}  B \Omega_\beta \| \Bigr\}
\cr
& \le \| A \| \, \| B \| = : M_{\Phi,\Psi} \qquad \forall z \in {\cal G}_{\delta /2}.}
For example,
\& {
\sup_{ \{ w \in \, \, \c : \Im w = \beta/2 \} }& f_{t^{(i)}_1,  \ldots, t^{(i)}_n} (w)   = \cr
& =
\sup_{s \in \r} \bigl( {\rm e}^{( i s + \beta / 2) H_\beta}  A \Omega_\beta \, , \,  
  F \bigl(t^{(i)}_1\bigr)    
\ldots F \bigl(t^{(i)}_n \bigr) {\rm e}^{i s H_\beta}  J^2 {\rm e}^{- \beta H_\beta
/ 2} 
 E B  \Omega_\beta \bigr)
\cr
& = \sup_{s \in \r} \bigl( {\rm e}^{  \beta H_\beta / 2}   A \Omega_\beta
 \, , \,   {\rm e}^{- i s H_\beta} F \bigl(t^{(i)}_1 \bigr)    
\ldots F \bigl(t^{(i)}_n \bigr) {\rm e}^{i s H_\beta} J  B^* E^*  \Omega_\beta \bigr)
\cr
& \le \| {\rm e}^{ \beta H_\beta / 2}   A \Omega_\beta \|  \| B \| 
\le \| J \| \| A^* \Omega_\beta \|  \| B \| \le \| A \|  \| B \|.}
Here we used $\| \Omega_\beta \| = 1$, $\| E \| = 1$, $\| F \| = 1$
and $\| {\rm e}^{- i s H_\beta} \| = 1$ for all $s \in \R$.
We emphasize that at this point also the specific properties of a KMS state are used; 
in the last line of (35) we made use of the modular conjugation
$J$ associated with the pair $({\cal R}_\beta, \Omega_\beta)$ (see, e.g.,
Ref.\ ( for a general account on modular theory).
By assumption $FE = 0$, hence
\# { f_{t^{(i)}_1, \ldots, t^{(i)}_n} \bigl( - t^{(i)}_j \bigr) = 0 .}
We conclude that inside the circle  $|z| < \delta / 2$ each of the
functions  $f_{t^{(i)}_1, \ldots, t^{(i)}_n} (z)$ posseses $n$ zeros for
pairwise different values of $t^{(i)}_j$. Thus all of the functions
\# { {f_{t^{(i)}_1, \ldots, t^{(i)}_n} (z)  \over \prod_{j = 1}^n (z + t^{(i)}_j)} ,
\qquad i \in \N,  }
are analytic in the open disk $D_{\delta / 2}$ of radius $\delta / 2 $ and
centered at the origin. Note that by definition 
$D_{ \delta / 2 } \subset {\cal G}_{ \delta / 2 }$.
Yet the number of zeros does not change in the limit $t^{(i)}_j \to 0$ and consequently,
for $i >1$,
\# { \Biggl| {  f_{t^{(i)}_1, \ldots, t^{(i)}_n} (z)  \over  
\prod_{j = 1}^n \bigl( z + t^{(i)}_j \bigr)  } \Biggr| 
\le
\sup_{w \in \partial D_{ \delta / 2 }} 
{  \bigl| f_{t^{(i)}_1, \ldots, t^{(i)}_n} (w) \bigr| \over  \prod_{j = 1}^n |w + t^{(i)}_j|  } 
\le M_{\Phi,\Psi} \cdot \Bigl({ 4  \over  \delta }\Bigr)^n  
\qquad \forall z \in D_{ \delta / 2 }.}
In the last inequality we used $\bigl| w + t_j^{(i)} \bigr| 
\ge \bigl| |w| - |t_j^{(i)}| \bigr|$ and
$|w| = \delta / 2 $ together with $\bigl| t_j^{(i)} \bigr| < \delta / 4$ for $i > 1$.
Hence,
\# {   \bigl| f_{ t^{(i)}_1, \ldots, t^{(i)}_n } (z)  \bigr|       
\le  M_{\Phi,\Psi} \cdot \Bigl({ 4  \over  \delta }\Bigr)^n   
\prod_{j = 1}^n \bigl|z + t^{(i)}_j \bigr|
\le  {\rm const} \cdot |z|^n  
\quad \hbox{as} \quad i \to \infty
\quad \forall z \in D_{ \delta / 2 }.}
Because of $F^2 = F$, it is obvious that 
$ f_{0, \ldots, 0} (z) = f_{E, F} (z)$.  
The map $t \mapsto {\rm e}^{it H_\beta}$ is  strongly continuous, thus
\# { \bigl| f_{t^{(i)}_1, \ldots, t^{(i)}_n} (z) - f_{0, \ldots, 0} (z)  \bigr| \to 0 
\qquad \hbox{as} \quad i \to \infty,}
uniformly in $z \in G_{ \delta / 2 }$. Thus
\# { \bigl| f_{E, F} (z) \bigr| \le  {\rm const} \cdot |z|^n  
\qquad \forall z \in D_{ \delta / 2 } .}
Hence $0$ is a zero of $n$th order. Since $n \in \N$ was arbitrary, we conclude that
$f_{E, F} (z)$ vanishes identically for all choices of
$\Phi , \Psi \in  {\cal D}_\tau $.  }

As a consequence of assumption (iii)\ $\omega_\beta$ is mixing and therefore the 
next lemma is more or less obvious.  

\Lm{Let $E, F \in \B(\H_\beta)$ be two projection operators and 
assume that 
\# { F {\rm e}^{i t H_\beta} E = 0 
\qquad \forall  t \in \R .}
It follows that $E \Omega_\beta \ne 0$ implies $F \Omega_\beta=0$.}

\Pr{By assumption, $\Omega_\beta$ is the unique --- up to a phase ---
normalized eigenvector for the discrete eigenvalue $\{ 0 \}$,
thus (Ref.\ 8)
\& { 0 & = \lim_{T \to \infty} {1 \over 2T} \int_{-T}^{T} {\rm d} t \, 
\bigl( \Omega_\beta \, , \, E {\rm e}^{it H_\beta} F \Omega_\beta \bigr)
\cr
& =  
(\Omega_\beta \, , \, E \Omega_\beta ) ( \Omega_\beta \, , F \Omega_\beta ) 
= \| E \Omega_\beta \|^2 \cdot \| F \Omega_\beta \|^2.}
If $E \Omega_\beta \ne 0$, it follows that  $F \Omega_\beta = 0$.}

We add a result whose analogon in the vacuum sector is due to Schlieder (see Ref.\ 18, p.\ 220).

\Cor{Let $E \in {\cal R}_\beta $  
be a nonzero projection  and $\xi \in \H_\beta$ an arbitrary nonzero vector. 
It follows that the set of points  
\# { \{ t \in \R : E {\rm e}^{i t H_\beta} \xi \ne 0 \} }
is dense in $\R$.}

\Pr{For $t \in \R$ fixed, the set of vectors 
\# { \H_t := \{ \Psi \in \H_\beta : E {\rm e}^{iH_\beta t} \Psi = 0 \} }
is a closed subspace of $\H_\beta$. We set
\# { \H_{] - \infty, \infty [} := \bigcap_{- \infty < t < \infty} \H_t .}
By construction, $\H_{] - \infty, \infty [}$ is the intersection of closed subspaces and therefore
$\H_{] - \infty, \infty [}$ itself is a  closed subspace of $\H_\beta$. 
Let $P$ denote the projection onto 
$\H_{] - \infty, \infty [}$.
Clearly,
\# {   E {\rm e}^{it H_\beta}  P  = 0 \qquad
\forall t \in \R. }
Now $0 \ne E \in {\cal R}_\beta$ implies $E \Omega_\beta \ne 0$. Therefore 
Lemma 2.2 implies $P \Omega_\beta = 0$. 
Since ${\rm e}^{-iH_\beta t} E {\rm e}^{iH_\beta t}  \in {\cal R}_\beta$ for all $t \in \R$, 
we conclude that
\# {  {\rm e}^{-iH_\beta t} E {\rm e}^{it H_\beta } D P = 0  \qquad
\forall D \in {\cal R}_\beta', \quad \forall t \in \R.}
Let $\hat {P} $ denote the projection 
onto the closed linear subspace
$  \overline { {\cal R}_\beta' \H_{ ]- \infty, \infty [} } $. 
Clearly,
\# {  E {\rm e}^{i t H_\beta }  \hat {P}  = 0   \qquad
\forall t \in \R . }
But by definition $ P $ is the maximal projection such that $E {\rm e}^{iH_\beta t}  P  = 0 $
holds for all $t \in \R$. It follows that $\hat {P} \le P$. On the other hand  
\# { \H_{] - \infty, \infty [} \subset 
\overline { {\cal R}_\beta' \H_{ ]- \infty, \infty [}} }
implies $P \le \hat {P}$, thus $P = \hat {P}$. We conclude
that $P \in {\cal R}_\beta$. Since $\Omega_\beta$ is separable for ${\cal R}_\beta$,
$P \Omega_\beta = 0$ implies $P= 0$.
It follows that for any vector $\xi \in \H_\beta$ there exists some $t \in \R$  such that
\# { E {\rm e}^{it H_\beta} \xi \ne 0 .}
Now consider the projection $P_\xi$ onto the one-dimensional subspace $\C \cdot \xi $ 
and assume there exists some $\delta >0$ such that 
\# { E {\rm e}^{i t H_\beta} P_\xi = 0 \qquad \forall |t-s| < \delta }
with $s \in \R$ fixed. Set 
$P_\xi (s) = {\rm e}^{ i s H_\beta}   P_\xi {\rm e}^{- i s H_\beta}$.  
Then $P_\xi (s) = P_\xi (s)^2 = P_\xi (s)^*$ is a projection and
\# {  E {\rm e}^{i t H_\beta}   P_\xi (s) = 0   \qquad
\forall |t| < \delta.  }
Lemma 2.1 implies that 
\# {  E {\rm e}^{i t H_\beta}   P_\xi = 0   \qquad
\forall t \in \R,  }
in contradiction to (51). Thus the set $ \{ t \in \R : E {\rm e}^{i t H_\beta} \xi \ne 0 \} $
does not contain any open intervall. Consequently, it is dense in $\R$.}

\Th{\hbox{(Schlieder property).}
Let $\O$ and $\hat {\O}$ denote two open (not necessarily bounded) 
space--time regions such that 
\# { \O + t e \subset \hat {\cal O}  
\qquad \forall |t| < \delta , \quad \delta >0.}
It follows that $0 \ne A \in {\cal R}_\beta (\O)$ and 
$0 \ne B \in {\cal R}_\beta (\hat {\O})$
implies $AB \ne 0$.}

\Pr{Let $A \in {\cal R}_\beta (\O) $ and
$B \in {\cal R}_\beta (\hat {\cal O})' $. We have to show that $AB  = 0$ implies $A = 0$ or $B=0$. 
From $AB  = 0$ we infer that both $A$ and $B$ can not be unitary. 
It follows that one of the expressions
$A^*A$ or $AA^*$ is unequal to $\1 $. The same is true for $B^*B$ or $BB^*$. 
Without loss of generality 
we assume that $A^*A \ne \1 $ and $BB^* \ne \1 $. With $A^*A$ also the spectral
projections of $A^*A$ belong to ${\cal R}_\beta (\O)$ and with $BB^*$ 
also the spectral projections of $BB^*$ belong to ${\cal R}_\beta (\hat {\cal O})'$. Thus
\# {A^*ABB^* = 0}
implies $FE = 0$ for all spectral projections $E \in {\cal R}_\beta (\O) $, 
$F \in {\cal R}_\beta (\hat {\cal O})' $ 
from the spectral resolution of 
$A^*A$ and $BB^*$, respectively. Since $E \in {\cal R}_\beta (\O) $ and 
$F \in {\cal R}_\beta (\hat {\cal O})' $ we find  
\# { \bigl[ {\rm e}^{- i t H_\beta} F {\rm e}^{ i t H_\beta} \, , \, E \bigr] = 0
\qquad \forall |t| < \delta.}
Onsequently, Lemma 2.1 implies
\# {  F {\rm e}^{i t H_\beta} E = 0
\qquad \forall t \in \R, }
and from Lemma 2.2 it follows that $E \Omega_\beta= 0$ or $F\Omega_\beta = 0$.
Finally, $\Omega_\beta$ is separating for both ${\cal R}_\beta (\O)$ and 
${\cal R}_\beta (\hat{\O})'$. Thus $E = 0$ or $F = 0$.}  

\Rem{The Schlieder property implies that ${\cal R}_\beta (\O)$ is almost a factor, namely
\# { {\cal R}_\beta (\O) \cap {\cal R}_\beta (\hat{\O})' = \C \cdot \1 .}
This can be seen as follows:
assume
\# { {\cal R}_\beta (\O) \cap {\cal R}_\beta (\hat{\O})' \ne \C \cdot \1 .}
It follows that there exists a nontrivial projection $P$ such that both
\# { P \in {\cal R}_\beta (\O) \cap {\cal R}_\beta (\hat{\O})'
\qquad \hbox{and} \qquad ( \1 - P ) \in {\cal R}_\beta (\O) \cap {\cal R}_\beta (\hat{\O})' .}
Set $A = P$ and $B = ( \1 - P ) $. The Schlieder property implies 
$P = 0$ or $\1 - P = 0$, in contradiction to the assumption that $P$ is 
a nontrivial projection.}

The Schlieder property is a first step towards
the ``statistical independence'' of ${\cal R}_\beta (\O)$ and
${\cal R}_\beta (\hat{\O})'$. In fact, several precise conditions 
for ``statistical independence''
have been proposed; an overview can be found in Ref.\ 19.
Florig and Summers have collected a list of properties which are equivalent to the  
Schlieder property.$^2$

\Cor{(Florig and Summers): Assume that $\H_\beta$ is separable.
Let $\O$, $\hat{\O}$ denote 
a pair of space--time regions such that the closure of the open 
region $\O$ is contained in the interior of $\hat{\O}$.
It follows that
\vskip .3cm
\noindent
\halign{ #  \hfil & \vtop { \parindent =0pt \hsize=36,6em
                            \strut # \strut} \cr 
(i)  & ${\cal R}_\beta (\O)$ and ${\cal R}_\beta (\hat{\O})'$ 
are $C^*$-independent, i.e., for every state $\omega_1$
on ${\cal R}_\beta (\O)$ and every state $\omega_2$ on 
${\cal R}_\beta (\hat{\O})'$ 
there exists a state $\omega$ on ${\cal R}_\beta$ 
such that $\omega_{| {\cal R}_\beta (\O)} = \omega_1 $ and 
$\omega_{| {\cal R}_\beta (\hat{\O})'} = \omega_2 $.
\cr
(ii)  & ${\cal R}_\beta (\O)$ and ${\cal R}_\beta (\hat{\O})'$ 
are $W^*$-independent, 
i.e., for every normal state $\omega_1$
on ${\cal R}_\beta (\O)$ and every normal state $\omega_2$ on 
${\cal R}_\beta (\hat{\O})'$ 
there exists a normal 
state $\omega$ on ${\cal R}_\beta$ 
such that $\omega_{| {\cal R}_\beta (\O)} = \omega_1 $ and 
$\omega_{| {\cal R}_\beta (\hat{\O})'} = \omega_2 $.
\cr
(iii)   &  For any nonzero vectors $\Phi$, $\Psi \in \H_\beta$ there 
exist $A' \in {\cal R}_\beta (\O)' $ and 
$B' \in {\cal R}_\beta (\hat{\O})$ such 
that  $A'\Phi = B'\Psi \ne 0$.
\cr
(iv)  & The ordered pair 
$\bigl( {\cal R}_\beta (\O),{\cal R}_\beta (\hat{\O})' \bigr)$ 
is strictly local; 
i.e., for any nonzero projection
$E \in {\cal R}_\beta (\O)$ and any state $\omega \in {\cal R}_\beta (\hat{\O})'_*$ there 
exists a state 
$\phi \in \bigl({\cal R}_\beta (\O) \vee {\cal R}_\beta (\hat{\O})'\bigr)_*$ 
such that 
$\phi (E) = 1$ and $\phi_{| {\cal R}_\beta (\hat{\O})'} = \omega$.
\cr
(v)  & For any nontrivial projections
$E \in {\cal R}_\beta (\O)$, $F \in {\cal R}_\beta (\hat{\O})'$ and $\lambda, \mu \in [0, 1]$ 
there exists a state 
$\phi$ on $\B(\H_\beta)$ such that $\phi (E) = \lambda$ and $\phi (F) = \mu$. 
\cr
(vi)   &  ${\cal R}_\beta (\O)$ and ${\cal R}_\beta (\hat{\O})'$ are 
statistically independent in the sense of Haag and Kastler;
i.e., for every state $\omega_1$
on ${\cal R}_\beta (\O)$ and every state $\omega_2$ on ${\cal R}_\beta (\hat{\O})'$ there 
exists a state $\omega$ on ${\cal R}$ 
such that 
\# { \omega (AB) = \omega_1 (A) \omega_2 (B) }
for all $A \in {\cal R}_\beta (\O)$ and all $B \in {\cal R}_\beta (\hat{\O})'$.
\cr
(vii)   &  $\| A B \| = \| A \| \,  \| B \|$
for all $A \in {\cal R}_\beta (\O)$ and all $B \in {\cal R}_\beta (\hat{\O})'$.
\cr
(viii)   &  The von Neumann algebras ${\cal R}_\beta (\O)$ and 
${\cal R}_\beta (\hat{\O})'$ 
are algebraically independent; i.e.,  
given  two arbitrary sets $\{ A_i : i = 1, \ldots, m \} $ 
and $\{ B_j : j = 1, \ldots, n \} $ 
of linear independent elements of ${\cal R}_\beta (\O)$ 
and ${\cal R}_\beta (\hat{\O})'$, 
respectively, 
the collection $\{ A_i B_j : i = 1, \ldots, m ; j = 1, \ldots, n \}$ 
is linearly 
independent in ${\cal R}_\beta (\O) \odot {\cal R}_\beta (\hat{\O})'$.
\cr
(ix)   &  The map $\eta \colon \bigl({\cal R}_\beta (\O) ,
{\cal R}_\beta (\hat{\O})' \bigr)
\to {\cal R}_\beta (\O) \odot {\cal R}_\beta (\hat{\O})'$ defined by
\# { \eta (AB) = A \otimes B, \qquad A \in {\cal R}_\beta (\O), 
\quad B \in {\cal R}_\beta (\hat{\O})',}
is an isomorphism continuous in the minimal $C^*$-cross norm on the algebraic tensor product 
${\cal R}_\beta (\O) \odot {\cal R}_\beta (\hat{\O})'$ and can therefore
be continuously extended to a surjective homomorphism $\overline{\eta} \colon
{\cal R}_\beta (\O) \vee {\cal R}_\beta (\hat{\O})' \to {\cal R}_\beta (\O) 
\otimes {\cal R}_\beta (\hat{\O})'$.
\cr}  
}

\Rem{If ${\cal R}_\beta (\O)$ is a factor of type 
III acting on a separable Hilbert space, 
then Corollary~2.5 remains valid, if we replace ${\cal R}_\beta (\hat{\O})'$
by ${\cal R}_\beta (\O)'$. It is remarkable that for such a pair all normal partial states 
have normal extensions, none of which is allowed to be a product state,
and also all partial states have extensions to product states, 
none of which can be normal.}

\Th{\hbox{(Borchers property).} Let $\O$ and $\hat {\O}$ denote two open and bounded
space--time regions such that 
\# { \O + t e \subset \hat {\cal O}  
\qquad \forall |t| < \delta , \quad \delta >0.}
Given a nonzero projection $E \in {\cal R}_\beta (\O)$, there exists a partial
isometry $V \in {\cal R}_\beta (\hat {\cal O})$ such that $V^* V = \1 $ and $V V^* = E$.}

\Pr{Once the Schlieder property is proven for ${\cal R}_\beta (\O)$
and ${\cal R}_\beta (\hat{\O})'$, the Borchers property follows by standard arguments
(see Ref.\ 1 and 16 for the corresponding result in the vacuum sector). 
We present them here for the sake of completeness only. 
By assumption the spacelike complement $\hat {\O}'$ of $\hat {\O}$
is not empty. Thus any vector
$\Phi \in {\cal D}_\tau$ is cyclic 
for ${\cal R}_\beta (\hat {\cal O})' \supset {\cal R}_\beta (\hat {\cal O}')$.
We show that $E \Phi$ is separating for ${\cal R}_\beta (\hat {\cal O})'$;
choose a region $\O_\circ$ such that
\# { \O_\circ \subset \O' \cap \hat {\cal O} }
and consider some $ B \in {\cal R}_\beta (\hat {\cal O})'$ such that $BE \Phi = 0$. 
Locality implies that $ BEC \Phi = 0$ for any $C \in {\cal R}_\beta (\O_\circ)$. 
By the Reeh-Schlieder property the set $\{ C \Phi : C \in {\cal R}_\beta (\O_\circ) \}$ is  
dense in $\H_\beta$ and therefore $BE = 0$.
Now the Schlieder property for ${\cal R}_\beta (\O)$ and
${\cal R}_\beta (\hat {\cal O})'$ implies $B = 0$, since by assumption $E \ne 0$. 
We conclude that $E \Phi$ is separating for ${\cal R}_\beta (\hat {\cal O})'$.
Hence, the normal state
\# {B \mapsto ( E \Phi \, , \, B E \Phi) }
is faithful on ${\cal R}_\beta (\hat {\cal O})'$ and 
there exists a vector $\Psi \in \H_\beta$ cyclic for 
${\cal R}_\beta (\hat {\cal O})'$ such that 
\# {( E \Phi \, , \, B E \Phi) = ( \Psi \, , \, B \Psi) 
\qquad \forall B \in {\cal R}_\beta (\hat {\cal O})'.}
It follows that $V \colon \H_\beta \to \H_\beta$, given by   
\# {V B \Psi = B E \Phi  \qquad \forall B \in {\cal R}_\beta (\hat {\cal O})',}
defines an isometry.
Both $\Phi$ and $\Psi$ are cyclic for ${\cal R}_\beta (\hat {\cal O})'$, thus
$V$ is densely defined and its range spans $E \H_\beta$. Moreover,
\# {C V B \Psi = C B E \Phi = V C  B \Psi \qquad \forall C \in {\cal R}_\beta (\hat {\cal O})'.}
Thus $V$ commutes with all $C \in {\cal R}_\beta (\hat {\cal O})'$ on the dense set  
$\{ B \Psi : B \in {\cal R}_\beta (\hat {\cal O})' \} \subset \H_\beta$ and
therefore
$V \in {\cal R}_\beta (\hat {\cal O})$.}

\Rem{The Borchers property has interesting consequences for the actual preparation of 
states: Given an arbitrary state $\omega$ on
${\cal R}_\beta (\O) \vee {\cal R}_\beta (\hat {\cal O})'$, we set
\# { \omega_V (C) := \omega (V^* C V )  
\qquad \forall  C \in {\cal R}_\beta (\O) \vee {\cal R} (\hat {\cal O})'.}
Then
\# { \omega_V (E) = \omega (V^* VV^* V ) = \1  \qquad \hbox{and} 
\qquad \omega_V (\1 - E) = 0 .}
Moreover,
\# { \omega_V (B) = \omega (V^* B V )  = \omega (B) \qquad 
\forall B \in {\cal R} (\hat {\cal O})'.}
This demonstrates that the Borchers property allows us to prepare a state $\omega_V$
which satisfies the properties (71) and (72) by a strictly local operation. 
We emphasize that the state given remains
completely unchanged in the  spatial complement of $\hat {\O}$. This is a remarkable 
difference to the usual collapse of the wave-function type of preparation.} 

\vskip .5cm

\noindent
{\it  Acknowledgements.\/}
\noindent
Kind hospitality of the Erwin Schr\"odinger Institut (ESI) Vienna and the
Dipartimento di Matematica, Universit\'a di Roma ``Tor Vergata'' is 
gratefully acknowleged. This work was financed --- at least in principle --- by
a fellowship of the Operator Algebras Network, EC TMR-Programme. 

\vskip .5cm

\noindent
{\fourteenrm References}
\nobreak
\vskip .3cm
\nobreak
\halign{   &  \vtop { \parindent=0pt \hsize=33em
                            \strut  # \strut} \cr 
\REF
{1}
{D'Antoni, C.}
                       {Technical properties of the quasi-local algebra}
                       {The algebraic theory of superselection sectors. Introduction and recent
                        results.  }{Proceedings of the convegno internationale algebraic 
                        theory of superselection sectors and field theory, Palermo
                        Nov.\ 23--30, 1989} 
                       {edited by} {D.\ Kastler}  
                       {World Scientific 1990}  
\REF
{2}
{Florig, S.\ and Summers, S.J.}      {On the statistical independence of algebras of observables}
                       {\JMP}
                       {38/3} {1318--1328}
                       {1997}
\BOOK
{3}
{Haag, R.}    {Local Quantum Physics: Fields, Particles, Algebras} 
              {Springer-Verlag, Berlin,} 
              {1992}
\REF
{4}
{Kubo, R.}    {Statistical mechanical theory of irreversible prozesses I.}    
              {J.\ Math.\ Soc.\ Jpn.}                                       
              {12} {570--586}                                       
              {1957}                                       
\REF
{5}
{Martin, P.C.\ and Schwinger, J.}   {Theory of many-particle systems.\ I}
                                    {Phys.\ Rev.}
                                    {115/6} {1342--1373}
                                    {1959}
\REF
{6}
{Haag, R., Hugenholtz, N.M.\ and Winnink, M.}
                          {On the equilibrium states in quantum statistical mechanics}  
                          {\CMP}	
                          {5}	{215--236}
                          {1967}
\REF
{7}
{Buchholz, D.\ and Junglas, P.}   {On the existence of equilibrium states in local 
                                   quantum field theory} 
                                  {\CMP} 
                                  {121} {255--270}
                                  {1989}
\BOOK
{8}  
{Bratteli, O.\ and Robinson, D.W.} {Operator Algebras and Quantum Statistical Mechanics~I, II} 
                                  {Sprin\-ger-Verlag, New York} 
                                  {1981}
\REF
{9}
{Haag, R., Kastler, D.\ and Trych-Pohlmeyer, E.B.}    {Stability and equilibrium states}
                                                      {\CMP} 
                                                      {38} {173--193}
						                                                {1974}
\REF
{10}
{Haag, R.\ and Trych-Pohlmeyer, E.B.}    {Stability properties of equilibrium states}
                                                      {\CMP} 
                                                      {56} {213--224}
						                                                {1977}
\REF
{11}
{Narnhofer, H.\ and Thirring, W.}   {Adiabatic theorem in quantum statistical mechanics}
                                   {Phys.\ Rev.}
                                   {A26/6}   {3646--3652} 
                                   {1982}
\REF
{12}
{Pusz, W., and Woronowicz, S.L.}   {Passive states and KMS states for general 
               quantum systems}
               {\CMP}
               {58}    {273--290}
               {1978}
\REF
{13}
{Bros, J.\ and Buchholz, D.}      {Towards a relativistic KMS-condition}
                                  {Nucl.\ Phys.\ B} 
                                  {429} {291--318}
                                  {1994}
\HEP
{14}
{J\"akel, C.D.}                    {The Reeh-Schlieder property for thermal field theories}
						                             {hep-th/9904049}
\BOOK
{15}
{Junglas, P.}   {Thermodynamisches Gleichgewicht und Energiespektrum in der 
                 Quantenfeldtheorie} 
                {dissertation, Hamburg}
                {1987}
\REF
{16}
{Borchers, H.J.}     {A remark on a theorem of B.\ Misra}
                    	{\CMP} 
                     {4}   {315--223}
                     {1967}
\BOOK
{17}
{Streater, R.F.\ and Wightman, A.S.}   {PCT, Spin and Statistics and all that}
                 {Benjamin, New York} 
                 {1964}
\REF
{18}
{Schlieder, S.}                     {Einige Bemerkungen \"uber Projektionsoperatoren}
                                    {\CMP}
                                    {13} {216--225}
                                    {1969}
\REF
{19}
{Summers, S.J.}       {On the statistical independence of algebras of observables}
                       {\RMP}
                       {2/2} {201--247}
                       {1990}
\cr}

\bye